\documentclass[letterpaper, 10 pt, conference]{ieeeconf}
\IEEEoverridecommandlockouts
\usepackage[T1]{fontenc}
\usepackage{balance}
\usepackage{silence}
\IEEEoverridecommandlockouts

\usepackage{cite}
\usepackage{graphicx}
\usepackage{picinpar}
\usepackage{stfloats}
\usepackage{url}
\usepackage{flushend}
\usepackage[latin1]{inputenc}
\usepackage{colortbl}
\usepackage{soul}
\usepackage{multirow}
\usepackage{pifont}
\usepackage{alltt}
\usepackage{enumerate}
\usepackage{siunitx}
\usepackage{pbox}
\usepackage{amsmath,amssymb,amsfonts}
\usepackage{amsthm}
\usepackage{algorithmic}
\usepackage{algorithm}
\usepackage{bm}
\usepackage{graphbox}
\usepackage{textcomp}
\usepackage{xcolor}
\usepackage[export]{adjustbox}
\usepackage{color}
\usepackage{subcaption}
\usepackage{tikz}
\def\BibTeX{{\rm B\kern-.05em{\sc i\kern-.025em b}\kern-.08em
    T\kern-.1667em\lower.7ex\hbox{E}\kern-.125emX}}

\definecolor{LightCyan}{rgb}{0.8,0.8,1.0}
\definecolor{LightRed}{rgb}{1.0,0.8,0.8}
\definecolor{LightGreen}{rgb}{0.8,1.0,0.8}
\definecolor{LightYellow}{rgb}{1.0,1.0,0.8}

\newcommand{\scaleMathLine}[2][1]{\resizebox{#1\linewidth}{!}{$\displaystyle{#2}$}}

\DeclareMathOperator{\diag}{diag}

\newtheorem{example}{Example}

\ActivateWarningFilters[balance]

\makeatletter
\let\NAT@parse\undefined
\makeatother
\usepackage[hidelinks]{hyperref}

\newcommand{\calD}{{\cal D}}
\newcommand{\calE}{{\cal E}}

\newcommand{\calL}{{\cal L}}

\newcommand{\calN}{{\cal N}}

\newcommand{\calV}{{\cal V}}

\newcommand{\bff}{\mathbf{f}}

\newcommand{\bfu}{\mathbf{u}}

\newcommand{\bfx}{\mathbf{x}}
\newcommand{\bfy}{\mathbf{y}}
\newcommand{\bfz}{\mathbf{z}}

\newcommand{\bftheta}{{\boldsymbol{\theta}}}

\newcommand{\bfpi}{{\boldsymbol{\pi}}}

\newcommand{\bfF}{\mathbf{F}}

\newcommand{\bfJ}{\mathbf{J}}

\newcommand{\bfR}{\mathbf{R}}

\newcommand{\bfZ}{\mathbf{Z}}

\newcommand{\bbR}{\mathbb{R}}

\newcommand{\prl}[1]{\left(#1\right)}
\newcommand{\brl}[1]{\left[#1\right]}

\title{\LARGE\bf LEMURS: Learning Distributed Multi-Robot Interactions}

\author{Eduardo Sebasti\'{a}n \and Thai Duong \and Nikolay Atanasov \and Eduardo Montijano \and Carlos Sag\"{u}\'{e}s%
\thanks{E. Sebasti\'{a}n, E. Montijano and C. Sag\"{u}\'{e}s are with the RoPeRt group, at DIIS - I3A, Universidad de Zaragoza, Spain (e-mails: \texttt{\small \{esebastian, emonti, csagues\}@unizar.es}).}%
\thanks{T. Duong and N. Atanasov are with the Department of Electrical and Computer Engineering, University of California San Diego, La Jolla, CA 92093 USA (e-mails: \texttt{\small \{tduong, natanasov\}@ucsd.edu}).}%
\thanks{This work has been supported by NSF CCF-2112665 (TILOS), by the ONR Global grant N62909-19-1-2027 and via Spanish projects PID2021-125514NB-I00, PID2021-124137OBI00 and TED2021-130224B-I00 funded by MCIN/AEI/10.13039/501100011033, by ERDF A way of making Europe and by the European Union NextGenerationEU/PRTR, DGA T45-20R, and Spanish grant FPU19-05700 and EST22/00253.}%
}

\newcommand\copyrighttext{%
  \footnotesize \textcopyright This paper has been accepted for publication in the 2023 IEEE International Conference on Robotics and Automation (IEEE ICRA 2023). Please cite the paper as: E. Sebasti\'{a}n, T. Duong, N. Atanasov, E. Montijano and C. Sag\"{u}\'{e}s,``LEMURS: Learning Distributed Multi-robot Interactions'', IEEE International Conference on Robotics and Automation (ICRA), 2023.}
\newcommand\copyrightnotice{%
\begin{tikzpicture}[remember picture,overlay]
\node[anchor=south,yshift=10pt] at (current page.south) {\fbox{\parbox{\dimexpr\textwidth-\fboxsep-\fboxrule\relax}{\copyrighttext}}};
\end{tikzpicture}%
}

\begin{document}
\maketitle
\copyrightnotice


\begin{abstract}
This paper presents LEMURS, an algorithm for learning scalable multi-robot control policies from cooperative task demonstrations. We propose a port-Hamiltonian description of the multi-robot system to exploit universal physical constraints in interconnected systems and achieve closed-loop stability. We represent a multi-robot control policy using an architecture that combines self-attention mechanisms and neural ordinary differential equations. The former handles time-varying communication in the robot team, while the latter respects the continuous-time robot dynamics. Our representation is distributed by construction, enabling the learned control policies to be deployed in robot teams of different sizes. We demonstrate that LEMURS can learn interactions and cooperative behaviors from demonstrations of multi-agent navigation and flocking tasks. 
\end{abstract}


\section{Introduction}\label{sec:intro}

Multi-robot systems promise improved efficiency and reliability compared to a single robot in many applications, including exploration and mapping \cite{atanasov2015decentralized,tian2022kimera}, agriculture and herding \cite{kan2021task,pierson2015bio,sebastian2021multi,sebastian2022adaptive}, and search and rescue \cite{williams2021search}. However, designing multi-robot control policies that achieve cooperative behaviors may be challenging. First, domain expertise may be required to specify the objective and constraints for a desired task in mathematical terms. Second, scaling the control policy to large teams may be computationally infeasible due to the increase of the joint state and control spaces. The first challenge motivates the use of machine learning techniques to learn reward functions or control policies from demonstration \cite{jiahao2021learning, bloembergen2015evolutionary, khan2020graph, tolstaya2020learning,tolstaya2021multi, yang2021communication, gama2022synthesizing, furieri2021distributed, han2020cooperative, shi2020neural, long2018towards, semnani2020multi, dasari2020robonet, bogert2018multi, zhu2021learning, zhou2019clone}. The second challenge motivates imposing a sparse structure in the control policy that respects the communication topology of the robot team and allows the complexity to scale with the number of neighbors \cite{khan2020graph, tolstaya2020learning, tolstaya2021multi,yang2021communication, gama2022synthesizing, furieri2021distributed, qu2020scalable, yang2018mean, wang2022darl1n}. In this work, we develop \textbf{LEMURS} (\textbf{LE}arning distributed \textbf{MU}lti-\textbf{R}obot interaction\textbf{S}), a learning approach for distributed control synthesis from cooperative task demonstrations that generalizes to different tasks, scales favorably with the number of robots, and handles time-varying robot communication.
 
Recent works focus on learning control policies for optimal control or reinforcement learning problems \cite{bloembergen2015evolutionary,long2018towards, semnani2020multi}. Given a cost function, a recurrent neural network \cite{zhu2021learning}, or graph convolutions and multi-layer perceptrons~\cite{zhou2019clone} have been used to learn centralized control policies. Without a cost function, inverse reinforcement learning \cite{ng2000algorithms} may be used to learn centralized \cite{dasari2020robonet,bogert2018multi} or distributed \cite{han2020cooperative} policies from task demonstrations. While black-box neural networks are widely used for learning control policies, they do not encode energy conservation and kinematic constraints satisfied by physical robot systems, and failing to infer them from data may result in unstable behaviors. A key contribution of our work is to represent the robot team as a \emph{port-Hamiltonian system} \cite{van2014port} and learn a distributed control policy from demonstration by modeling robot interactions as energy exchanges. The use of Hamiltonian mechanics has been explored for centralized control policies or fixed-time known topologies~\cite{furieri2021distributed, galimberti2021hamiltonian}, in which scalability is achieved in the absence of communication \cite{shi2020neural}. Meanwhile, our work achieves scalability with a time-varying topology by modeling robot interactions using \emph{self-attention techniques}~\cite{vaswani2017attention}.

Learning and execution of control policies for multi-robot systems should scale favorably with an increasing numbers of robots. Learning a joint value or policy function is challenging due to the exponential growth of the state and action space \cite{qu2020scalable}. Successful methods for multi-agent reinforcement learning factorize value functions  according to the $k$-hop neighborhoods \cite{wang2022darl1n,qu2020scalable} or using attention mechanisms \cite{long2020evolutionary}. Graph neural networks have been utilized as a scalable and communication-aware policy representation in coverage, exploration, and flocking problems \cite{khan2020graph, tolstaya2020learning,tolstaya2021multi,yang2021communication, gama2022synthesizing}. Recently, Li et al.~\cite{li2021message} combine graph-neural networks with self-attention to solve decentralized multi-robot path planning problems. Many of these techniques assume discrete robot dynamics, fixed or known communication topology, or prior knowledge about the task. In contrast, our approach learns from demonstrated robot trajectories with an unknown task objective and handles time-varying communication and team sizes. 
In order to handle continuous-time dynamics, we use \emph{neural ordinary differential equation (ODE) networks} \cite{chen2018neural}.
Closely related, Jiahao et al. \cite{jiahao2021learning} develop a neural ODE network that learns distributed controllers but enforces collision avoidance using an explicit potential field and assumes a fixed maximum number of neighbors. By using a port-Hamiltonian formulation and self-attention mechanism, we handle time-varying neighbors, do not constrain the size of the neighborhoods, and learn constraints such as collision avoidance from data.

In summary, we develop LEMURS, a novel algorithm for learning scalable multi-robot control policies from demonstration. Our \emph{first contribution} is the use of port-Hamiltonian dynamics to restrict the family of learned policies to those that are stable and distributed. Our \emph{second contribution} is a novel learning architecture that integrates concepts of self-attention and neural ODEs to handle continuous-time dynamics, time-varying communication, and large robot teams. 


\section{Problem Statement}\label{sec:prosta}

Consider a team of robots, indexed by $\mathcal{V} = \{1, \hdots, n \}$. Assume that the dynamics of each robot $i \in \mathcal{V}$ are \emph{known}:
\begin{equation}\label{eq:known_dynamics}
    \dot{\mathbf{x}}_i(t) = \bff_i(\mathbf{x}_i(t),\mathbf{u}_i(t)),
\end{equation}
%
where $\mathbf{x}_i(t) \in \mathbb{R}^{n_x}$ and $\mathbf{u}_i(t) \in \mathbb{R}^{n_u}$ denote the state and control input of the robot at time $t\geq0$. The robots interact in a distributed manner, described by a time-varying undirected graph $\mathcal{G}(t) = (\mathcal{V},\mathcal{E}(t))$, where $\mathcal{E}(t) \subseteq \mathcal{V} \times \mathcal{V}$ is the set of edges. An edge $(i,j) \in \mathcal{E}(t)$ exists when robots $i$ and $j$ interact at time $t$. Robot $i$ can always interact with itself, i.e., $(i,i) \in \calE(t)$ for all $i$, $t$. Let $\mathbf{A}(t) \in \{0,1\}^{n \times n}$ be the weighted adjacency matrix associated to $\mathcal{G}(t)$, such that $[\mathbf{A}(t)]_{ij} \neq 0$ if and only if $(i,j) \in \mathcal{E}(t)$, and $0$ otherwise. The set of $k$-hop neighbors of robot $i$ at $t$ is $\mathcal{N}_i^k(t) = \{j \in \mathcal{V} \mid [\mathbf{A}^k(t)]_{ij} \neq 0 \}$. 
Each robot executes an \emph{unknown} control policy:
\begin{equation}\label{eq:general_controller}
    \mathbf{u}_i(t) = \bfpi_{\bftheta}\left(\bfx_{\calN_i^k}(t)\right),
\end{equation}
where $\bfx_{\calN_i^k}(t) = \{\bfx_j(t) \mid j \in \calN_i^k(t)\}$ and $\bftheta$ is the control policy parameters. Our objective is to use task demonstrations to learn $\bftheta$, such that the multi-robot team, following the learned control policy, replicates the task. 

\begin{example}\label{example:flocking_controller}
Consider a flocking task \cite{olfati2006flocking} in which a robot team must achieve a formation with aligned velocities, while avoiding collisions. The robots follow double integrator dynamics $\dot{\mathbf{p}}_i(t) = \mathbf{v}_i(t)$ and $\dot{\mathbf{v}}_i(t) = \mathbf{u}_i(t)$, where $\mathbf{p}_i(t) \in \mathbb{R}^m$, $\mathbf{v}_i(t) \in \mathbb{R}^m$, and $\mathbf{u}_i(t) \in \mathbb{R}^m$ are the position, velocity, and input of robot~$i$. A distributed control policy that achieves flocking was developed by Olfati-Saber \cite{olfati2006flocking}:
\begin{align}
    \mathbf{u}_i(t) =& - c_1\mathbf{p}_i(t) - c_2\mathbf{v}_i(t) + \kern -0.3cm \sum_{j \in \mathcal{N}^1_i(t)} \kern -0.3cm \phi(||\Delta\mathbf{p}_{ij}(t)||_{\sigma})\mathbf{n}_{ij}(t) 
    \notag\\&
    + \kern -0.2cm \sum_{j \in \mathcal{N}^1_i(t)}\kern -0.2cm \rho\left(||\Delta\mathbf{p}_{ij}(t)||_{\sigma}\right)(\mathbf{v}_{j}(t)-\mathbf{v}_i(t))
     \label{eq:real_flocking_controller}
\end{align}
where $\Vert{\cdot}\Vert_{\sigma}$ is the $\sigma$-norm of a vector and $\Delta\mathbf{p}_{ij}(t) = \mathbf{p}_{j}(t)-\mathbf{p}_i(t)$.
The first and second terms are a proportional controller with gains $c_1,c_2>0$ that prevents the formation of sub-flocks, where we have assumed $\mathbf{p}_i(t) = \mathbf{v}_i(t) = \mathbf{0}$ $\forall i$ as the desired flock configuration. The third term avoids robot collisions and induces the desired lattice formation, where $\phi(\cdot)$ is a potential function whose minima are located at the desired inter-robot distances, and $\mathbf{n}_{ij}(t)$ is a vector that points in the repulsion/coalition direction.  The last term achieves velocity consensus using a distance scaling function $\rho(\cdot)$ that models the robot communication. Further details can be found in \cite{olfati2006flocking}. This paper aims to learn policies like \eqref{eq:real_flocking_controller} from demonstrations.
\end{example}



We assume that state trajectories from successful task executions are available as training data. Let $\mathbf{x}(t) = [\mathbf{x}_1^\top(t), \ldots, \mathbf{x}_n^\top(t)]^\top$ and $\mathbf{u}(t) = [\mathbf{u}_1^\top(t), \ldots, \mathbf{u}_n^\top(t)]^\top$ denote the joint state and control of the robot team. Given an initial state $\mathbf{x}(0) = \mathbf{x}_0$, let $\bar{\mathbf{x}}_{0:K} := [\bar{\mathbf{x}}(0), \hdots, \bar{\mathbf{x}}(rT), \hdots, \bar{\mathbf{x}}(KT)]$ and $\mathbf{x}_{0:K} := [\mathbf{x}(0), \hdots, \mathbf{x}(rT), \hdots, \mathbf{x}(KT)]$ be demonstrated and learned policy's trajectories, respectively, with $K$ denoting the number of discrete samples $r$ along the trajectories with sampling interval $T$. Let $\bar{\mathcal{D}} := \{ \bar{\mathbf{x}}_{0:K}^l \}_{l=0}^{l=L}$ denote a dataset of $L >0$ demonstrated trajectories. Let $\mathcal{D} := \{ \mathbf{x}_{0:K}^l \}_{l=0}^{l=L}$ be the generated trajectories under policy $\pi_{\bftheta}$. We aim to learn a control policy that minimizes the distance between the demonstrated and generated trajectories:
\begin{equation}\label{eq:loss_function}
    \mathcal{L}({\mathcal{D}},\bar{\mathcal{D}}) = \frac{1}{KL}\sum_{l=0}^L\sum_{r=0}^K
    ||\mathbf{x}^l(rT)-\bar{\mathbf{x}}^l(rT)||^2_2.
\end{equation}
Formally, the problem we consider is:
\begin{subequations}\label{eq:prob_def}
\begin{alignat}{2}
\min_{\bftheta} & \hbox{ }       \mathcal{L}({\mathcal{D}},\bar{\mathcal{D}}) \label{eq:prob_def_cost}
\\
\text{s.t.} & \hbox{ }\dot{\mathbf{x}}_{i}^l(t) = \bff_i(\mathbf{x}_i^l(t), \mathbf{u}_i^l(t)), \hbox{ } \bfx_i^l(0) = \bar{\bfx}_i^l(0), \hbox{ } \forall i,l,\label{eq:prob_def_constraint2}
\\
& \hbox{ } {\mathbf{u}}_i^l(t) = \bfpi_{\bftheta}(\bfx^l_{\calN_i^{k}}(t)), \hbox{ } \forall i,l,t. \label{eq:prob_def_constraint3}
\end{alignat}
\end{subequations}
As specified in the formulation above, the learned control policy should also handle time-varying communication and should be adaptable to changes in the total number of robots or the number of neighbors for each robot.





\section{LEMURS}\label{sec:solution}

In this section, we present a port-Hamiltonian formulation of the multi-robot dynamics and an energy-based distributed control design that can shape the interactions and Hamiltonian of the closed-loop system (Sec.~\ref{subsec:portHamiltonian}). Given task demonstrations, we employ self-attention and neural ordinary differential equations to learn the interactions and energy parameters of the control policy that minimize the distance between the demonstrated and generated trajectories (Sec. \ref{subsec:LEMURS}). To simplify the notation, we omit the time dependence of the states $\bfx$ and controls $\bfu$ in the remainder of the paper.



\subsection{Port-Hamiltonian Formulation of Multi-Robot Dynamics}\label{subsec:portHamiltonian}

Port-Hamiltonian mechanics are a general yet interpretable modeling approach for learning and control. On the one hand, many physical networked systems can be described as a port-Hamiltonian system~\cite{furieri2021distributed} using the same formulation and with a modular and distributed interpretation. Meanwhile, the port-Hamiltonian description allows to derive general energy-based controllers with closed-loop stability guarantees. Since robots are physical systems that satisfy Hamiltonian mechanics, we model each robot's dynamics in \eqref{eq:known_dynamics} as a port-Hamiltonian system \cite{van2014port}:
\begin{equation}\label{eq:open_loop_intro_individual}
        \dot{\mathbf{x}}_i = \left(\mathbf{J}_s^{(i)}(\mathbf{x}_i) - \mathbf{R}_s^{(i)}(\mathbf{x}_i)\right) \frac{\partial {H}_s^{(i)}(\mathbf{x}_i)}{\partial\mathbf{x}_i}
        + \mathbf{F}_s^{(i)}(\mathbf{x}_i)\mathbf{u}_i,
\end{equation}
where the skew-symmetric interconnection matrix $\mathbf{J}_s^{(i)}(\bfx_i)$ represents energy exchange \emph{within a robot}, the positive-semidefinite dissipation matrix $\bfR_s^{(i)}(\bfx_i)$ represents energy dissipation, the Hamiltonian ${H}_s^{(i)}(\mathbf{x}_i)$ represents the total energy, and the matrix $\bfF_s^{(i)}(\bfx_i)$ is the input gain. Then, the multi-robot system with joint state $\bfx$ also follows port-Hamiltonian dynamics:
\begin{equation}\label{eq:open_loop_intro}
        \dot{\mathbf{x}} =  \left(\mathbf{J}_s(\mathbf{x})  -  \mathbf{R}_s(\mathbf{x})\right)\frac{\partial {H}_s(\mathbf{x})}{\partial\mathbf{x}}  +  \mathbf{F}_s(\mathbf{x})\mathbf{u},
\end{equation}
where $H_s(\bfx) = \sum_{i=1}^n H_s^{(i)}(\bfx_i)$ and
\begin{equation}\label{eq:block_diag_structure}
\begin{aligned}
    \bfJ_s(\bfx) &= \text{diag}\left(\bfJ_s^{(1)}(\bfx_1), \ldots, \bfJ_s^{(n)}(\bfx_n) \right), \\
    \bfR_s(\bfx) &= \text{diag}\left(\bfR_s^{(1)}(\bfx_1), \ldots, \bfR_s^{(n)}(\bfx_n) \right), \\
    \bfF_s(\bfx) &= \text{diag}\left(\bfF_s^{(1)}(\bfx_1), \ldots, \bfF_s^{(n)}(\bfx_n) \right).
\end{aligned}
\end{equation}
Without control, the trajectories of the open-loop system in \eqref{eq:open_loop_intro} would not match the demonstrations in $\bar{\calD}$. The dynamics need to be controlled by the policy in \eqref{eq:general_controller} in order to generate desired trajectories. We employ an interconnection and damping assignment passivity-based control (IDA-PBC) approach \cite{van2014port}, which injects additional energy to the system through the control input $\bfu$ to achieve some closed-loop dynamics that replicate the demonstrated task:
\begin{equation}\label{eq:closed_loop_intro}
        \dot{\mathbf{x}} = \left( \mathbf{J}_{\bftheta}(\mathbf{x})   -   \mathbf{R}_{\bftheta}(\mathbf{x})\right)
\frac{\partial {H_{\bftheta}}(\mathbf{x})}{\partial\mathbf{x}}, 
\end{equation}
with Hamiltonian $H_{\bftheta}(\bfx)$, skew-symmetric interconnection $\mathbf{J}_{\bftheta}(\mathbf{x})$, and positive semidefinite dissipation $\mathbf{R}_{\bftheta}(\mathbf{x})$. By matching the terms in \eqref{eq:open_loop_intro} and \eqref{eq:closed_loop_intro}, one obtains the policy:
\begin{equation}\label{eq:open_loop_intro_controller}
\begin{aligned}
\mathbf{u} &= \mathbf{F}_s^{\dagger}(\mathbf{x})
\left( \left(\mathbf{J}_{\bftheta}(\mathbf{x}) - \mathbf{R}_{\bftheta}(\mathbf{x})\right)\frac{\partial {H_{\bftheta}}(\mathbf{x})}{\partial\mathbf{x}} \right. 
\\
& \qquad \quad \kern 0.2cm- \left.\left(\mathbf{J}_s(\mathbf{x}) - \mathbf{R}_s(\mathbf{x})\right)\frac{\partial {H}_s(\mathbf{x})}{\partial\mathbf{x}} \right),
\end{aligned}
\end{equation}
where $\mathbf{F}_s^{\dagger}(\mathbf{x}) = \left(\bfF_s^{\top}(\bfx)\bfF_s(\bfx)\right)^{-1}\bfF_s^{\top}(\bfx)$ is the pseudo-inverse of $\mathbf{F}_s(\mathbf{x})$. 
If the robots are fully-actuated, i.e., $\bfF_s(\bfx)$ is full-rank, the matching condition on the pseudo-inverse is always satisfied, achieving the desired closed-loop dynamics. For underactuated systems, satisfaction of the matching condition may not always be possible \cite{blankenstein2002matching}. Being able to achieve zero error $\calL(\calD,\bar{\calD})$ is, hence, related to whether the demonstrated trajectories $\bar{\calD}$ are realizable by the class of control policies in \eqref{eq:open_loop_intro_controller}. Even if the trajectories in $\bar{\calD}$ are not realizable, the policy parameters $\bftheta$ may still be optimized to achieve a behavior as similar as possible.

Let $[\mathbf{J}_{\bftheta}(\mathbf{x})]_{ij}$ and $[\mathbf{R}_{\bftheta}(\mathbf{x})]_{ij}$ denote the $n_x \times n_x$ blocks with index $(i,j)$, representing the energy exchange between robot $i$ and $j$ and the energy dissipation of robot $i$ caused by robot $j$, respectively. Since the input gain $\mathbf{F}_s(\mathbf{x})$ in \eqref{eq:block_diag_structure} is block-diagonal, the individual control policy of robot $i$ is:
\begin{equation}\label{eq:open_loop_intro_controller_ind}
\begin{aligned}
\kern -5pt \mathbf{u}_i &= \mathbf{F}_s^{(i)\dagger}(\mathbf{x}_i)
\left( \sum_{j \in \calV}\left([\mathbf{J}_{\bftheta}(\mathbf{x})]_{ij} - [\mathbf{R}_{\bftheta}(\mathbf{x})]_{ij}\right)\frac{\partial H_{\bftheta}(\mathbf{x})}{\partial \mathbf{x}_j}\right. 
\\
& \qquad \quad \kern 0.3cm- \left.\left(\mathbf{J}_s^{(i)}(\mathbf{x}_i) - \mathbf{R}_s^{(i)}(\mathbf{x}_i)\right)\frac{\partial {H}_s^{(i)}(\mathbf{x})}{\partial\mathbf{x}_i} \right).
\end{aligned}
\end{equation}
The individual control policies in \eqref{eq:open_loop_intro_controller_ind} do not necessarily respect the hops in the communication network as desired in \eqref{eq:general_controller} because this depends on the structure of $\mathbf{J}_{\bftheta}(\mathbf{x})$,  $\mathbf{R}_{\bftheta}(\mathbf{x})$, and $H_{\bftheta}(\bfx)$. In Sec.~\ref{subsec:LEMURS}, we impose conditions on these terms to ensure that they respect the communication topology and are skew-symmetric, and positive semidefinite, respectively, as required for a valid port-Hamiltonian system.

\begin{example}\label{example:portHamiltonian}
By substituting the control policy~\eqref{eq:real_flocking_controller} in the double integrator dynamics of the robots, the closed-loop dynamics for the flocking problem in Example~\ref{example:flocking_controller} are:
\begin{equation}\label{eq:closed_loop_new}
\begin{aligned}
        \dot{{\mathbf{p}}} &= {\mathbf{v}},\quad
        \dot{{\mathbf{v}}} = -\frac{\partial {U}_{\bftheta}({\mathbf{p}})}{\partial{\mathbf{p}}} - \mathbf{D}_{\bftheta}({\mathbf{p}}){\mathbf{v}}, \\\hbox{ with }
        \frac{\partial {U}_{\bftheta}({\mathbf{p}})}{\partial{\mathbf{p}}_i}  &= c_1{\mathbf{p}}_i + \sum_{j \in \mathcal{N}_i^1}  \phi(||{\mathbf{p}}_{j}-{\mathbf{p}}_i||_{\sigma})\mathbf{n}_{ij},
        \\
         [\mathbf{D}_{\bftheta}({\mathbf{p}})]_{ij}  &=  \left(  c_2 [\mathbf{I}_n]_{ij}  +  \rho \left(||{\mathbf{p}}_{j}-{\mathbf{p}}_i||_{\sigma}/\tau\right) \right) \mathbf{I}_m.
\end{aligned}
\end{equation}
In port-Hamiltonian terms, ${H}_{s}(\bfx) = \frac{1}{2}{\mathbf{v}}^\top{\mathbf{v}}$, $\bfR_s = \mathbf{0}$, $\mathbf{F}_s(\mathbf{x}) = [\mathbf{0}, \mathbf{I}_m]^\top$, 
$\bfJ_{s}(\bfx) = \bfJ_{\bftheta}(\bfx) =          \begin{pmatrix}
        \mathbf{0} & \kern -0.2cm \mathbf{I}_n\\ -\mathbf{I}_n & \kern -0.2cm\mathbf{0}
        \end{pmatrix}$, $\bfR_{\bftheta}(\bfx) =          \begin{pmatrix}
        \mathbf{0} &  \mathbf{0}\\ \mathbf{0} &  \mathbf{D}_{\bftheta}({\mathbf{p}})
        \end{pmatrix}$, and ${H}_{\bftheta}(\bfx) = U_{\bftheta}({\mathbf{p}}) + \frac{1}{2}{\mathbf{v}}^\top{\mathbf{v}}$.
\end{example}

\subsection{Learning Distributed Multi-Robot Interactions}\label{subsec:LEMURS}

The analytical design of scalable cooperative control policies like the flocking controller of Example~\ref{example:flocking_controller} is challenging when the complexity of the task increases. Instead, we seek to learn control policies that scale with the number of robots, handle time-varying communications and guarantee the port-Hamiltonian constraints. To do so, we first derive conditions on $\mathbf{J}_{\bftheta}(\mathbf{x})$, $\mathbf{R}_{\bftheta}(\mathbf{x})$ and ${H}_{\bftheta}(\mathbf{x})$. Then, we develop a novel architecture based on self-attention and neural ordinary differential equations to ensure that the learned control policies guarantee these conditions.

We first impose  $\mathbf{J}_{\bftheta}(\mathbf{x})$ and $\mathbf{R}_{\bftheta}(\mathbf{x})$ to be block-sparse, 
\begin{equation} 
\label{eq:JRH_conditions1}
        [\mathbf{J}_{\bftheta}(\mathbf{x})]_{ij} =
        [\mathbf{R}_{\bftheta}(\mathbf{x})]_{ij} = \mathbf{0}, \quad \forall j \notin \calN_i^k.
\end{equation}
This is to satisfy the topology constraints of the multi-robot team. Moreover, 
we require that the desired Hamiltonian factorizes over $k$-hop neighborhoods:
\begin{equation} \label{eq:individual_Hi}
        H_{\bftheta}(\mathbf{x}) = \sum_{i=0}^n H_{\bftheta}^{(i)}(\bfx_{\calN_i^{k}}).
\end{equation}
The factorization in \eqref{eq:individual_Hi} ensures that each robot $i$ can calculate  $\partial H_{\bftheta}(\bfx)/\partial \mathbf{x}_i = \sum_{j\in \calN_i^k}{\partial H^{(j)}_{\bftheta}(\bfx_{\calN_j^{k}})/\partial \mathbf{x}_i}$ by gathering $\partial H^{(j)}_{\bftheta}(\bfx_{\calN_j^{k}})/\partial \mathbf{x}_i$ from its $k$-hop neighbors $j$. 
Then, the control policy $\bfpi_{\bftheta}$ of robot $i$ becomes:
\begin{align}
\kern -5pt \mathbf{u}_i &= \mathbf{F}_s^{(i)\dagger}(\mathbf{x}_i)
\left( \sum_{j \in \calN_i^k}\kern -0.2cm \left([\mathbf{J}_{\bftheta}(\mathbf{x})]_{ij} \kern -0.1cm-\kern -0.1cm [\mathbf{R}_{\bftheta}(\mathbf{x})]_{ij}\right)\frac{\partial H_{\bftheta}(\mathbf{x})}{\partial \mathbf{x}_j}  \right. 
\notag\\
& \qquad \qquad \left. -\left(\mathbf{J}_s^{(i)}(\mathbf{x}_i) - \mathbf{R}_s^{(i)}(\mathbf{x}_i)\right)\frac{\partial {H}_s^{(i)}(\mathbf{x})}{\partial\mathbf{x}_i} \right). \label{eq:open_loop_intro_controller_ind_distributed}
\end{align}
Imposing the requirements in \eqref{eq:JRH_conditions1}-\eqref{eq:individual_Hi} is a first step towards making the control policy in \eqref{eq:open_loop_intro_controller_ind_distributed} distributed. Note that the terms $[\mathbf{J}_{\bftheta}(\mathbf{x})]_{ij}$ and $[\mathbf{R}_{\bftheta}(\mathbf{x})]_{ij}$ might still depend on the joint state $\bfx$ even if $j \in \calN_i^k$. We discuss how to remove this dependence next and achieve a similar factorization as \eqref{eq:individual_Hi}.

\subsubsection{Modeling robot interactions using self-attention}
\label{subsubsec:sa_architecture_JRH}
We model $[\mathbf{J}_{\bftheta}]_{ij}, [\mathbf{R}_{\bftheta}]_{ij}$, and $H_{\bftheta}^{(i)}$ in
Eq.~\eqref{eq:open_loop_intro_controller_ind_distributed} with the parameters $\bftheta$ shared across the robots, so that the team can handle time-varying communication graphs.
Specifically, we propose a novel architecture 
based on self-attention~\cite{vaswani2017attention}. Self-attention consists of a sequence of operations (a layer) that extracts the relationships among the inputs of a sequence by calculating the importance associated to each input using an attention map. The length of the sequences can vary as the number of parameters of the self-attention is constant with the number of inputs. Our key idea is to consider the self and neighboring information as the sequence, where each neighbor's information is an input. 

\begin{figure}[t]
    \centering
    \includegraphics[width=\linewidth]{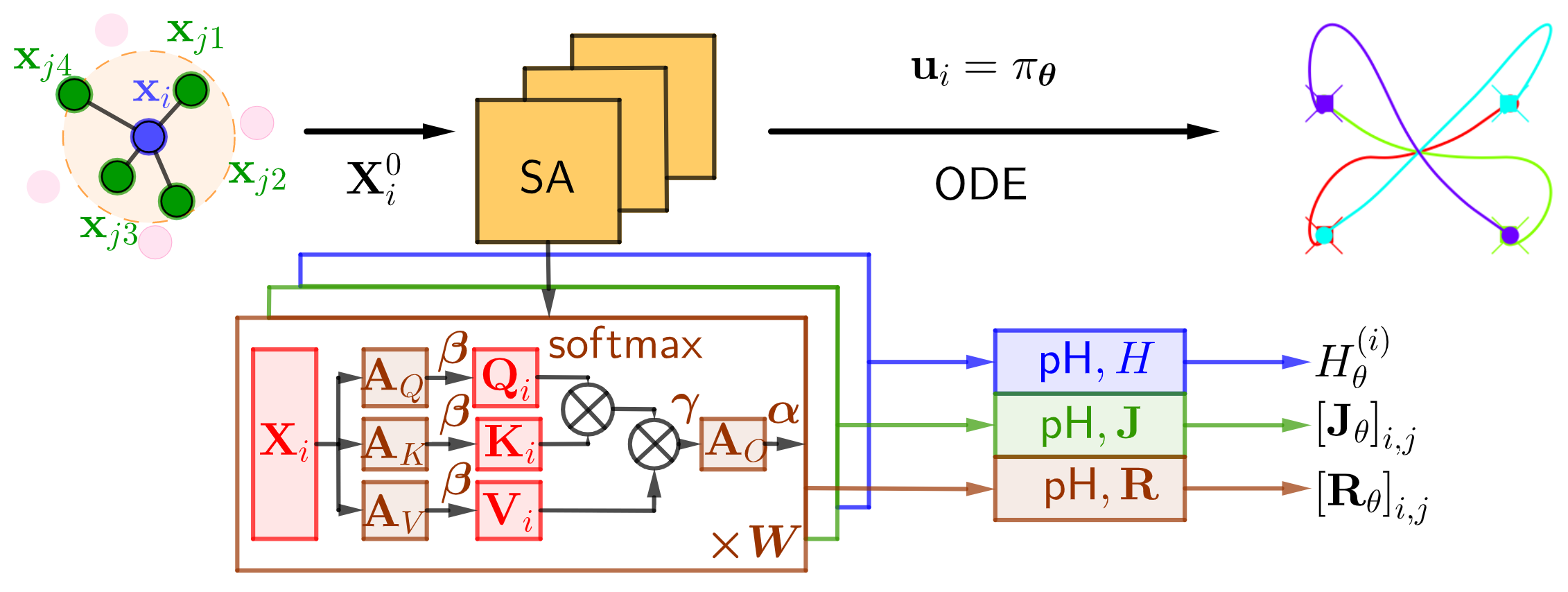}
    \caption{\small{Architecture of LEMURS: robot $i$ receives information from its neighbors. Then, the self-attention module obtains the port-Hamiltonian terms. Finally, it computes the control policy through an ordinary differential equation solver.}}
    \label{fig:architecture}
\end{figure}

To learn
$[\mathbf{R}_{\bftheta}]_{ij},$
robot $i$ will use, at instant $t$, the states $\bfx_j$ from all $k-$hop neighbors $j \in \mathcal{N}^k_i$, concatenated as follows:
\begin{equation} \label{eq:concatenated_neighbors}
    \mathbf{X}_i^0 = [\mathbf{x}_{i},     \mathbf{x}_{j_1}, \mathbf{x}_{j_2}, \hdots, \mathbf{x}_{j_{\left|\mathcal{N}_i^k\right|}}] \in \bbR^{n_x \times (|\calN_i^k| + 1)}.
\end{equation}
For each layer $w=1,\hdots, W$, we define:
\begin{align}\label{eq:selfattention}
        &\kern -0.2cm \mathbf{Q}_i^w =  \mathbf{A}^w_{Q,\mathbf{R}}{\mathbf{X}}_i^w, \hbox{ } \mathbf{K}_i^w = \mathbf{A}^w_{K,\mathbf{R} }\mathbf{X}_i^w, \hbox{ } \mathbf{V}_i^w = \mathbf{A}^w_{V,\mathbf{R}}\mathbf{X}_i^w, 
    \\
    &\kern -0.2cm\mathbf{Y}_i^w = \gamma\left(\mathsf{softmax}\left(\frac{\beta(\mathbf{Q}^w_i)\beta((\mathbf{K}_i^w)^{\top})}{\sqrt{|\mathcal{N}_i|}}\right)\beta(\mathbf{V}_i^w)\right),\label{eq:selfattention1}   
    \\
    &\kern -0.2cm\mathbf{X}_i^w = \alpha(\mathbf{A}^w_{Z, \mathbf{R}}{\mathbf{Y}}_i^w),\label{eq:selfattention2}   
\end{align}
where \textsf{softmax} stands for the softmax operation; $W$ is the number of self-attention layers; $\mathbf{A}^w_{Q, \mathbf{R}}, \mathbf{A}^w_{K, \mathbf{R}}, \mathbf{A}^w_{V, \mathbf{R}} \in \mathbb{R}^{r_w \times h_w}$ and $\mathbf{A}^w_{Z, \mathbf{R}} \in \mathbb{R}^{d_w \times r_w}$ for $w =1, \ldots, W$ are matrices to be learned and shared across robots; and $h_w, r_w, d_w>0$, with $d_W = n_x^2$ and $h_1 = n_x$ for valid matrix multiplications. The size of $\mathbf{A}^w_{Q, \mathbf{R}}, \mathbf{A}^w_{K, \mathbf{R}}, \mathbf{A}^w_{V, \mathbf{R}}$ does not depend on the number of robots, so robot $i$ can deal with time-varying neighbors. Nonlinear activation functions $\beta(\cdot)$, $\gamma(\cdot)$ and $\alpha(\cdot)$ account for potential nonlinearities. The concatenation in~\eqref{eq:concatenated_neighbors} is valid since the self-attention equation~\eqref{eq:selfattention1} learns the relationship among all the elements of $\mathbf{X}^w_i$ via the inner matrix multiplication.
Then, $[\mathbf{R}_{\bftheta}]_{ij}$ is constructed as a weighted matrix that models the interactions of robot $i$ with its neighbors, and a diagonal positive semidefinite matrix that accounts for the self-interactions:
\begin{gather}\label{eq:Restimate}
    \begin{aligned}
    \mathbf{Z}_{ij}^R =& \mathsf{vec}^{-1}(\mathbf{x}_{ij}^W)
    \\
   [\mathbf{R}_{\bftheta}]_{ij} =& -({\mathbf{Z}}_{ij}^{R}+{\mathbf{Z}}_{ji}^{R}), \quad  \forall j \in \mathcal{N}^k_i\backslash \{i\},
    \\
    [\mathbf{R}_{\bftheta}]_{ii} =& {\mathbf{Z}}_{ii}^{R} +  \sum_{j \in \mathcal{N}_i^k \backslash \{i\}}  ({\mathbf{Z}}_{ij}^{R}+{\mathbf{Z}}_{ji}^{R}),
    \end{aligned}
\end{gather}
where $\mathbf{x}_{ij}^W$ is the column $\mathbf{x}_j$ that corresponds to neighbor $j$ in $\mathbf{X}_i^W$,  $\mathsf{vec}^{-1}(\cdot)$ is the operator that reshapes the $d_W \times 1$ vector to a $n_x \times n_x$ matrix.
This way, $[\mathbf{R}_{\bftheta}]_{ij}$ is positive semidefinite by design.


To construct $[\mathbf{J}_{\bftheta}]_{ij}$, we follow the same steps \eqref{eq:selfattention}-\eqref{eq:selfattention2}, with parameters $\mathbf{A}^w_{Q, \mathbf{J}}, \mathbf{A}^w_{K, \mathbf{J}}, \mathbf{A}^w_{V, \mathbf{J}}$ and $\mathbf{A}^w_{Z, \mathbf{J}}$, to obtain encodings $\bfZ_{ij}^J$. Due to the undirected communication between robots $i$ and $j$, we enforce the skew-symmetry of $\bfJ_\bftheta$ by:
\begin{equation}\label{eq:Jestimate}
    [\mathbf{J}_{\bftheta}]_{ij} = {\mathbf{Z}}_{ij}^{J}-{\mathbf{Z}}_{ji}^{J} \quad \forall j \in \mathcal{N}^k_i.
\end{equation}

For each robot $i$, we construct $H^{(i)}_{\bftheta}$ as follows:
\begin{equation}\label{eq:Hestimate}
H^{(i)}_{\bftheta} = (\mathsf{vec}(\mathbf{X}_i^0))^{\top} \mathbf{M}_{\bftheta}^{(i)}(\mathbf{X}_i^0)(\mathsf{vec}(\mathbf{X}_i^0)) + {U}_{\bftheta}^{(i)}(\mathbf{X}_i^0),
\end{equation}
where the first term $(\mathsf{vec}(\mathbf{X}_i^0))^{\top} \mathbf{M}_{\bftheta}^{(i)}(\mathbf{X}_i^0)(\mathsf{vec}(\mathbf{X}_i^0))$ is a kinetic-like energy function with $\mathbf{M}_{\bftheta}^{(i)}(\mathbf{X}_i^0) = \diag( \mathbf{1}^{\top}\mathbf{Z}^M_i)$, and the second term ${U}_{\bftheta}^{(i)}(\mathbf{X}_i^0)$ is a potential energy function with ${U}_{\bftheta}^{(i)}(\mathbf{X}_i^0) = \mathbf{1}^{\top}\mathbf{Z}^U_i\mathbf{1}$.
The encodings $\mathbf{Z}^M_i$ and $\mathbf{Z}^U_i$ are calculated using the same steps ~\eqref{eq:selfattention}-\eqref{eq:selfattention2}, with parameters  $\mathbf{A}^w_{Q, \mathbf{M}}, \mathbf{A}^w_{K, \mathbf{M}}, \mathbf{A}^w_{V, \mathbf{M}}$ and $\mathbf{A}^w_{Z, \mathbf{M}}$, and $\mathbf{A}^w_{Q, U}, \mathbf{A}^w_{K, U}, \mathbf{A}^w_{V, U}$ and $\mathbf{A}^w_{U}$, respectively.
With $H^{(i)}_{\bftheta}$, we  obtain $\frac{\partial H^{(i)}_{\bftheta}}{\partial \mathbf{x}_j}$   $\forall j \in \mathcal{N}^k_{i}$ and compute  $\frac{\partial H_{\bftheta}}{\partial \bfx_i} = \sum_{j \in \mathcal{N}_i^k } \frac{\partial H^{(j)}}{\partial \bfx_i}$. 


\subsubsection{Learning distributed control policies using neural ODE networks}
Let
$\mathsf{SA}(\mathbf{X}_i^0,\bm{\theta})$ be the operations~\eqref{eq:selfattention}-\eqref{eq:Hestimate} with $$\bm{\theta} = \{ \{\mathbf{A}^w_{Q, k}, \mathbf{A}^w_{K, k}, \mathbf{A}^w_{V, k}, \mathbf{A}^w_{Z, k}\}_{w=1}^{w=W}\}_{k=\{\mathbf{R,J,M},U\}}.$$
To address Problem \eqref{eq:prob_def}, we use a neural ODE network \cite{chen2018neural} whose structure respects the continuous-time dynamics in \eqref{eq:open_loop_intro_individual}. To calculate the loss $\mathcal{L}({\mathcal{D}},\bar{\mathcal{D}})$ in \eqref{eq:loss_function}, for each trajectory $l$ of robot $i$, $\{\bfx_i^l(rT)\}_{r = 0}^K$ in the data, we solve an ODE:
\begin{equation}
\dot{\bfx}^l_i = \bff_i(\bfx^l_i, \bfpi_\bftheta;\bftheta), \quad \bfx^l_i(0) = \bar{\bfx}^l_i(0),
\end{equation}
using an ODE solver to obtain a predicted state $\bfx_i^l(rT)$ for $i \in \calV, l = 0, \ldots, L$: 
\begin{equation}
\bfx^l_i = \mathsf{ODESolver}\prl{\bfx^l_i(0), \bff_i, rT; \bftheta}.
\end{equation}
The parameters $\bftheta$ are updated using gradient descent by back-propagating the loss through the neural ODE solver using adjoint states $\bfy_i = \frac{\partial \calL}{\partial {\bfx_i}}$ \cite{chen2018neural}. We form an augmented state $\bfz_i = \prl{\bfx_i, \bfy_i, \frac{\partial \calL}{\partial \bftheta}}$ that satisfies $\dot{\bfz}_i = \bff_\bfz = \prl{\bff_i, -\bfy_i^\top \frac{\partial \bff_i}{\partial {\bfx_i}}, -\bfy_i^\top \frac{\partial \bff_i}{\partial \bftheta}}$. The gradients $\frac{\partial \calL}{\partial \bftheta}$ are obtained by solving a reverse-time ODE starting from $\bfz_i(rT) = \bar{\bfz}_i(rT)$:
\begin{equation}
\prl{\bfx_i(0), \bfy_i(0), \partial \calL/\partial\bftheta} = \mathsf{ODESolver}(\bfz_i(rT), \bff_z, rT).
\end{equation}
We refer the reader to \cite{chen2018neural} for more details.

\subsubsection{Deploying LEMURS}
\label{subsec:msg}
To deploy the control policy \eqref{eq:open_loop_intro_controller_ind_distributed}, we design a message $\mathbf{m}_{ij}(t)$, encoding information that robot $i$ needs from robot $j$ at time $t$ to calculate $[\mathbf{J}_{\bftheta}]_{ij}, [\mathbf{R}_{\bftheta}]_{ij}$, and $H_{\bftheta}^{(i)}$. 

Each robot $i$ receives a message $\mathbf{m}_{ij} = [\mathbf{m}^{(1)}_{ij}, \mathbf{m}^{(2)}_{ij}, \mathbf{m}^{(3)}_{ij}]$ in $3$ communication rounds: 1) robot $i$ receives $\mathbf{m}^{(1)}_{ij} = \bfx_j$ $\forall j \in \calN_i^k$ and calculates ${\mathbf{Z}}_{ij}^{J}$, ${\mathbf{Z}}_{ij}^{R}$, $H_{\bftheta}^{(i)}$, and $\partial H^{(i)}_{\bftheta}/\partial \mathbf{x}_j$; 2) robot $i$ receives $\mathbf{m}^{(2)}_{ij} = \partial H^{(j)}_{\bftheta}/\partial \mathbf{x}_i, {\mathbf{Z}}_{ji}^{J}, {\mathbf{Z}}_{ij}^{R}$ $\forall j \in \calN_i^k$, and calculates $\partial H_{\bftheta}/\partial \mathbf{x}_i$, $[\mathbf{J}_{\bftheta}]_{ij}, [\mathbf{R}_{\bftheta}]_{ij}$; and 3) each robot $i$ receives $\mathbf{m}^{(3)}_{ij} = \partial H_{\bftheta}/\partial \mathbf{x}_j$ $\forall j \in \calN_i^k$ and calculates the control input $\bfu_i$.
We assume negligible delays between communication rounds. If the delay is large, Wang et al. ~\cite{wang2022darl1n} suggest to learn a function that predicts quantities such as $\partial H_{\bftheta}(\mathbf{x}) / \partial \mathbf{x}_j$, ${\mathbf{Z}}_{ji}^{J}$, ${\mathbf{Z}}_{ji}^{R}$, leading to one communication round. We leave this for future work. If the Hamiltonian changes slowly over sampling interval $T$, at time $t = rT$, robot $i$ can use its previous neighbor states $\bfx_j ((r-1)T)$ to approximate $\mathbf{m}^{(2)}_{ij}(rT)$ and $\mathbf{m}^{(3)}_{ij}(rT)$. 



\begin{example}\label{example:flockingsolution}
In the flocking of Examples~\ref{example:flocking_controller}-\ref{example:portHamiltonian}, LEMURS can be directly applied to learn $\mathbf{D}_{\bftheta}(\mathbf{p})$ and ${U}_{\bftheta}(\mathbf{p})$. Another option is to learn $\mathbf{J}_{\bftheta}(\mathbf{x}), \mathbf{R}_{\bftheta}(\mathbf{x})$ and ${H}_{\bftheta}(\mathbf{x})$, obtaining extra degrees of freedom for the control policy.
\end{example}


\section{Results}\label{sec:simulations}

\begin{figure}[!ht]
    \centering\hspace{-0.6cm}
    \begin{tabular}{cc}
    \subcaptionbox{Fixed swapping, ECP\label{fig:fixedswapping_gtc}}
      {\includegraphics[width=0.48\columnwidth]{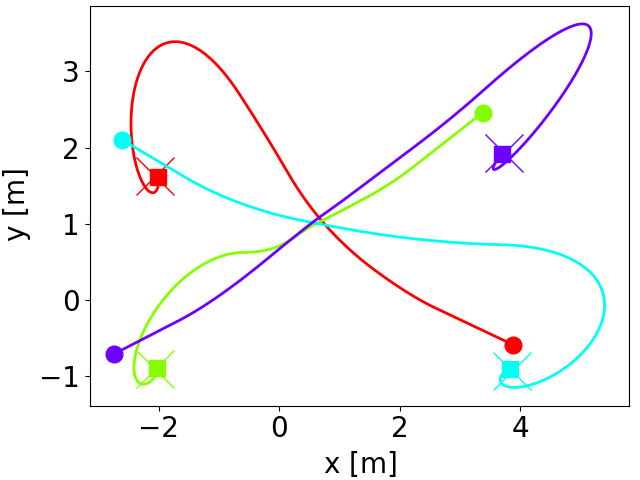}}
      &  
    \subcaptionbox{Fixed swapping, LCP\label{fig:fixedswapping_lcp}}
      {\includegraphics[width=0.48\columnwidth]{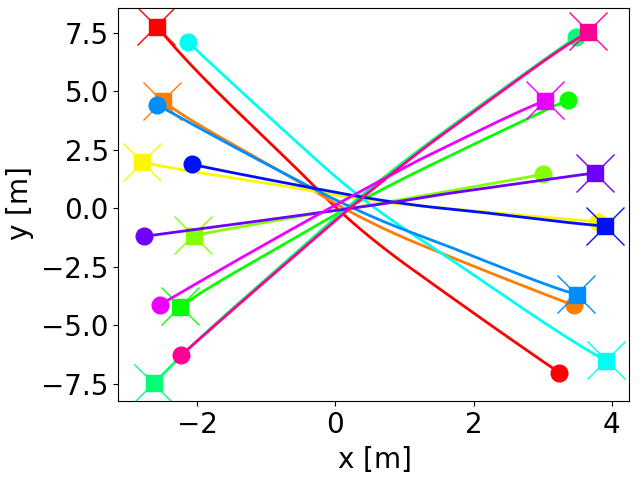}}
      \\
    \subcaptionbox{TV swapping, ECP\label{fig:tvswapping_gtc}}
      {\includegraphics[width=0.48\columnwidth]{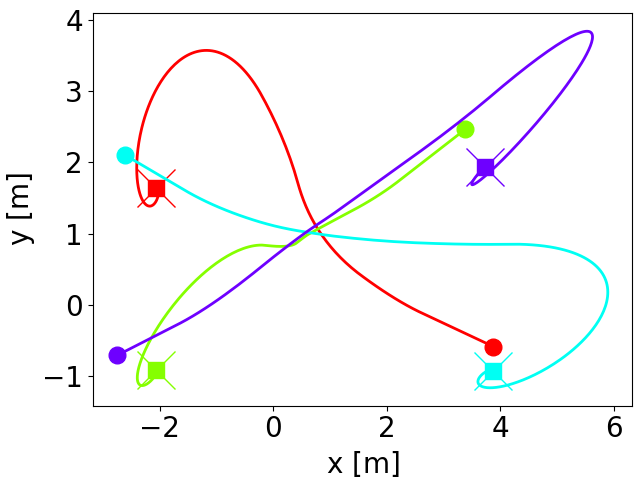}}
      &
    \subcaptionbox{TV swapping, LCP\label{fig:tvswapping_lcp}}
      {\includegraphics[width=0.48\columnwidth]{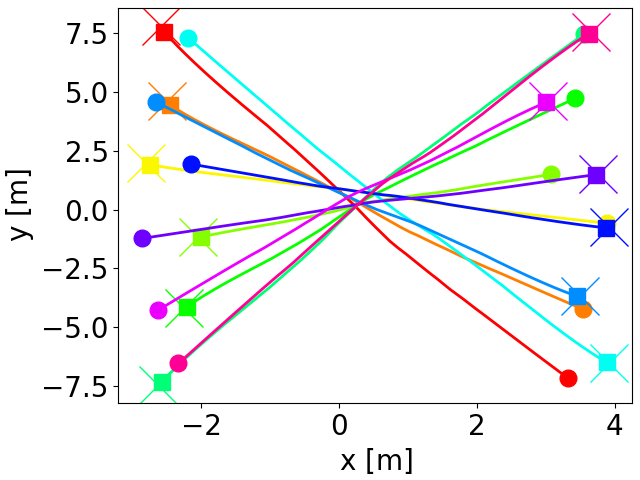}}
      \\
    \subcaptionbox{Flocking, ECP\label{fig:flocking_gtc}}
      {\includegraphics[width=0.48\columnwidth]{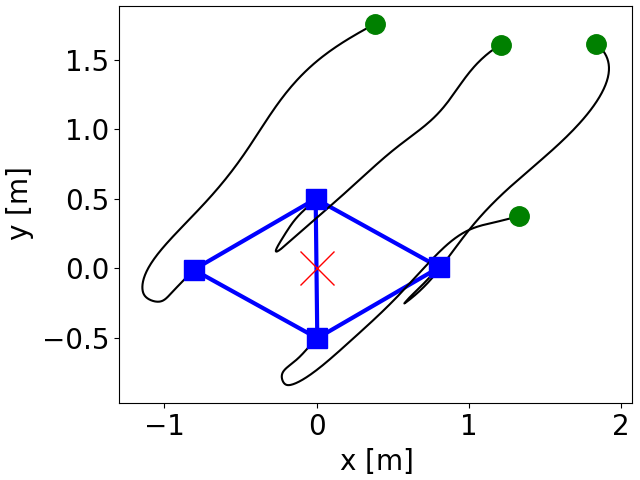}}
      &
    \subcaptionbox{Flocking, LCP\label{fig:flocking_lcp}}
      {\includegraphics[width=0.48\columnwidth]{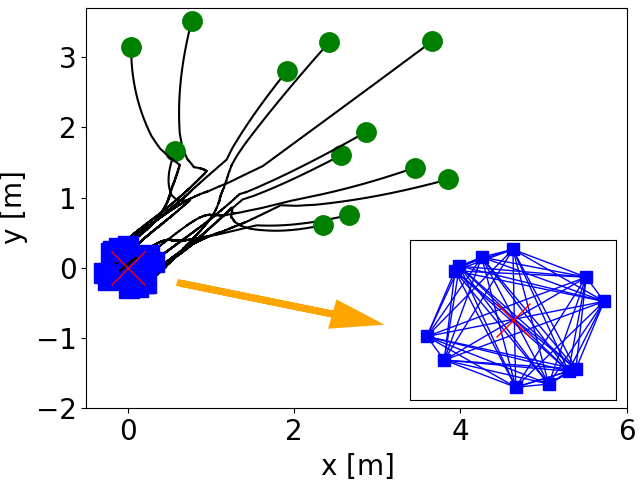}}
    \end{tabular}
	\caption{\small Demonstration of expert and learned control policies: (left) expert control policies (ECP) for training with $4$ robots, (right) learned control policies (LCP) with $12$ robots for $3$ tasks.
	}
	\label{fig:examples}
\end{figure}

In this section we evaluate LEMURS in three multi-robot tasks with simulated point robots, illustrated in Fig.~\ref{fig:examples}:
\begin{enumerate}
    \item \emph{Fixed swapping} \cite{furieri2021distributed}: Robots are initialized in two columns and navigate to the diagonally opposite position in the other column while avoiding collisions (Fig.~\ref{fig:fixedswapping_gtc}). The communication graph is a fixed ring such that robot $i$ communicates with robots $(i\pm1)\mod n$. We use the same parameters as~\cite{furieri2021distributed}. We generate demonstrations from the following expert controller:
    \begin{equation} \label{eq:swapping-expert}
            \mathbf{u}_i = -c_1\mathbf{p}_i-c_2\mathbf{v}_i - \sum_{j\in\mathcal{N}_i^1}\frac{\mathbf{p}_{i}-\mathbf{p}_{j}}{\sqrt{1+\sigma||\mathbf{p}_{i}-\mathbf{p}_{j}||^2_2}},
    \end{equation} with $c_1=0.8$, $c_2=1.0$, $\sigma = 0.1$.
    
    \item \textit{Time-varying swapping}: We consider the \textit{fixed swapping} task but with a time-varying communication graph (Fig.~\ref{fig:tvswapping_gtc}), where $\mathbf{A}(t)$ is such that $[\mathbf{A}(t)]_{ij} = \mathsf{sigmoid}(\lambda(||\mathbf{p}_i(t)-\mathbf{p}_j(t)|| - l/2))$ if $||\mathbf{p}_i(t)-\mathbf{p}_j(t)||<l$ and $0$ otherwise, with $l = 2.4$m and $\lambda = 2.0$. We use the controller in \eqref{eq:swapping-expert} with time-varying neighbors to generate demonstrations.
    
    \item \textit{Flocking}: We consider the flocking task described in Examples~\ref{example:flocking_controller}-\ref{example:flockingsolution} with parameters from \cite{olfati2006flocking} (Fig.~\ref{fig:flocking_gtc}). We use the controller \eqref{eq:real_flocking_controller} to generate demonstrations. 
\end{enumerate}


The training and evaluation datasets for each task have $L=400$ trajectories of $K=250$ samples with sampling interval $T=0.04$s. The number of demonstrating robots is $n=4$, and the trajectories are split in sub-trajectories of $5$ samples for training. We train for $10000$ epochs with learning rate $0.001$, and new batches of $200$ samples every $100$ epochs. The \textsf{ODEsolver} is the Euler numerical method \cite{butcher2016numerical}. We consider $k = 1$ as the number of hops.

The learned control policies are stable and scalable for all the tasks, converging to the desired goal with a larger number of robots as seen in Figs.~\ref{fig:examples}. We plot trajectories from the expert control policy (left) and learned control policies for $12$ robots (right), three times larger than the team size in training. Similar results with up to $64$ robots can be found on our website\footnote{\scriptsize \url{https://eduardosebastianrodriguez.github.io/LEMURS/}}.  For all three tasks, LEMURS achieves similar performance compared to analytical policies, which were used to generate training trajectories. LEMURS successfully captures behaviors that are not encoded a priori in the architecture of the networks nor in the cost function, such as the collision avoidance or the flock formation in \textit{flocking}. Collision avoidance among the point robots is verified in all tasks by checking the distance between each pair of robots.  In the \textit{swapping} problems, as the training dataset is formed by sub-trajectories of $5$ samples, which resemble a straight line in general, LEMURS infers that the motion to the goals should be a straight line as well. On the other hand, the minimum distance among robots is $0.005$m, avoiding collisions even in the center of the stage. In flocking task, since we train LEMURS for \emph{flocking} with only $4$ robots in Fig. \ref{fig:flocking_gtc}, LEMURS infers that it is desired to have $4$ groups of robots with equal distances between the groups (Fig. \ref{fig:flocking_lcp}), prioritizing formation to safety. In this sense, evaluation with $4$ robots yields to a minimum distance among robots of $0.05$m, while with $12$ robots the minimum distance among robots is $0.001$m. To improve generalization, we suggest increasing the number of robots during training, but we leave this for future work. 

We compare LEMURS with three other learning methods: 1) Multi-Layer Perceptron (MLP), inspired by~\cite{furieri2021distributed}; 2) Graph Neural Network (GNN) from~\cite{khan2020graph,yang2021communication}; and 3) Self-Attention based Graph Neural Network (GNNSA)~\cite{li2021message}, which uses graph neural networks preceded by a self-attention layer to model communication channels. These learning models substitute the self-attention layers in our Hamiltonian-based neural ODE networks. We keep the port-Hamiltonian neural ODE architecture for a fair comparison with the other discrete-time and/or black-box policies, leaving the complete adaptation of the other papers to our setting for future work. The size of the layers/filters in the MLP, GNN and GNNSA depends on the number of robots, so scalability is not directly achievable unlike in our approach. LEMURS has $2208$ parameters while MLP, GNN and GNNSA have $1$ layer/filter with $4448, 4448$ and $4672$ parameters, respectively. The details are in the Appendix~\ref{sec:parameters}. 

\begin{figure}[!ht]
    \centering\hspace{-0.6cm}
    \begin{tabular}{cc}
    \subcaptionbox{Fixed swapping,evaluation }
      {\includegraphics[width=0.45\columnwidth]{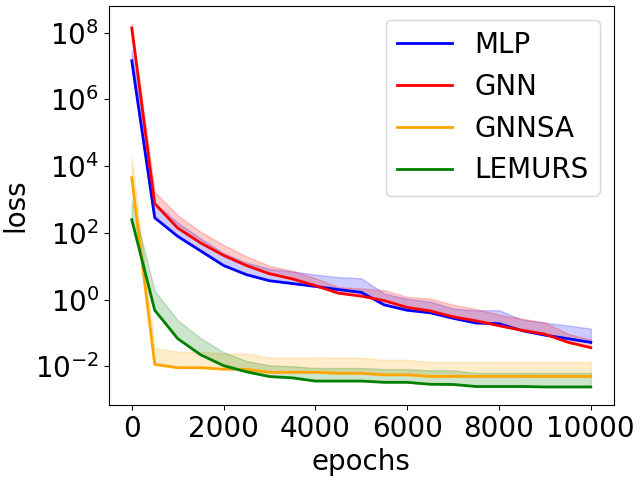}}
      &  
    \subcaptionbox{Fixed swapping,scalability}
      {\includegraphics[width=0.45\columnwidth]{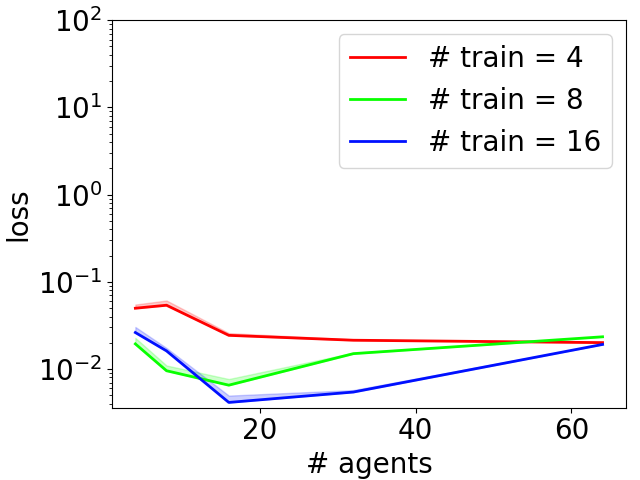}}
      \\
    \subcaptionbox{TV swapping, evaluation }
      {\includegraphics[width=0.45\columnwidth]{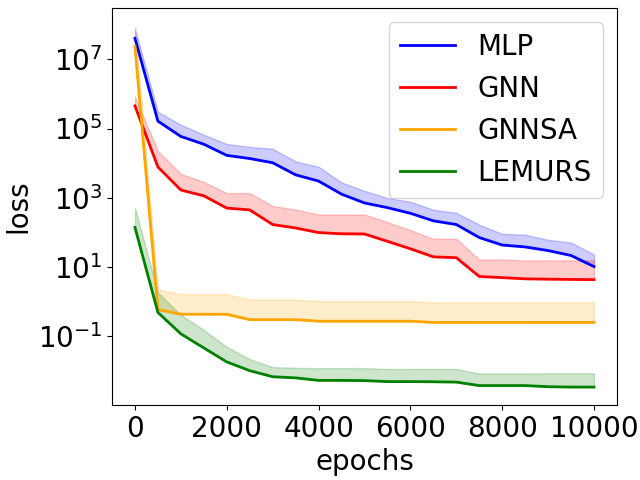}}
      &
    \subcaptionbox{TV swapping, scalability}
      {\includegraphics[width=0.45\columnwidth]{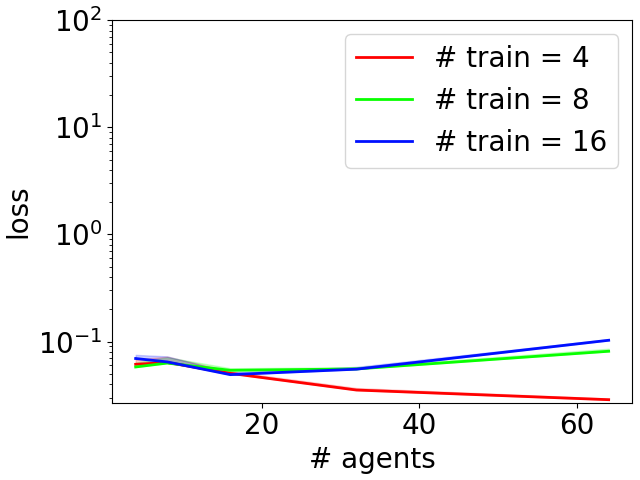}}
      \\
    \subcaptionbox{Flocking, evaluation \label{fig:training_flocking}}
      {\includegraphics[width=0.45\columnwidth]{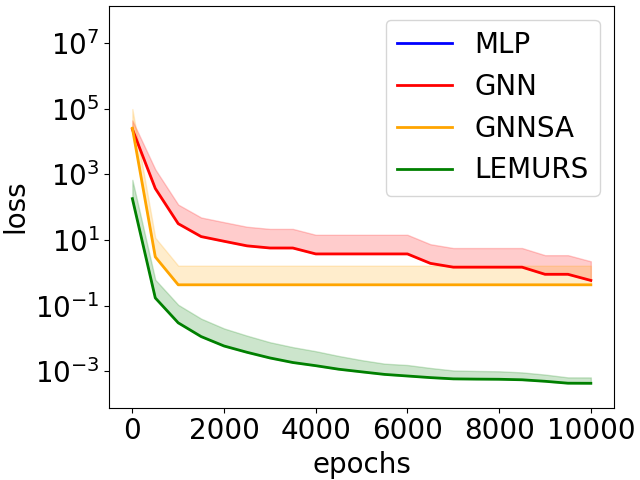}}
      &
    \subcaptionbox{Flocking, scalability}
      {\includegraphics[width=0.45\columnwidth]{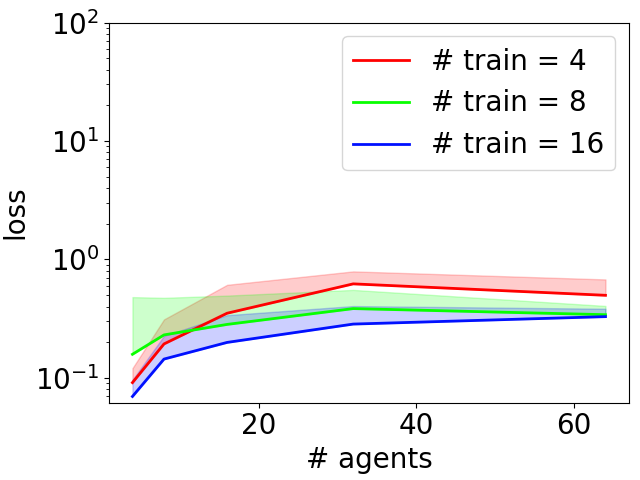}}
    \end{tabular}
	\caption{\small Comparison of LEMURS with a multi-layer perceptron (MLP)~\cite{furieri2021distributed}, a graph neural network (GNN)~\cite{khan2020graph,yang2021communication} and a self-attention-based graph neural network (GNNSA)~\cite{li2021message} in learning robot interactions: (left) evaluation loss over $10000$ epochs, (right) LEMURS training scalability with $4, 8, 16$ robots for $3$ tasks.
	}
	\label{fig:training_and_scalability}
\end{figure}

Fig.~\ref{fig:training_and_scalability} (left) plots the evaluation loss of the $4$ models and the $3$ tasks, from $3$ runs using $3$ randomized seeds. Our self-attention architecture surpasses the other three methods in capturing interactions for all tasks with half of the number of parameters, illustrating the benefits of combining self-attention networks and Hamiltonian architecture in LEMURS. In our experiments, self-attention is shown to learn more complex aggregation patterns compared to graph neural networks, potentially because in graph neural networks the data is aggregated via a pre-multiplication of a linear graph shift operator, while SA aggregates data through Eq.~\eqref{eq:selfattention1}. LEMURS achieves training loss two orders of magnitudes better than that of MLP, GNN and GNNSA in tasks with time-varying topologies. Meanwhile, the MLP training did not converge with data from the \emph{flocking} task. For the \textit{fixed swapping} task, LEMURS's evaluation loss is slightly better than that of GNNSA, and two orders of magnitudes better than that of MLP and GNN. 




We also validate scalability. The policies are simulated over a time horizon $KT = 10$s. We train LEMURS with datasets of $n=\{4, 8, 16\}$, for $5$ runs using randomized seeds, and test the learned control policies with $n=\{4, 8, 16, 32, 64\}$. The mean and standard deviation of the test loss (Eq.~\eqref{eq:loss_function}) is normalized by $n$ and plotted in Fig.~\ref{fig:training_and_scalability} (right). LEMURS obtains similar test loss with respect to the number of training robots. In the case of \textit{fixed swapping}, increasing the number of robots in training improves the controller performance since the larger number of robots is, the more data is available to learn about a fixed communication topology.
For the \textit{time-varying swapping} task, a small number of robots in training performs slightly better, potentially because the time-varying topology is more complex with more robots. Meanwhile, for \textit{flocking} task, increasing the number of training robots slightly improves the performance, even though the topology is also time-varying. This is because the robots form a flocking formation in the training trajectories, leading to a fixed topology in a large portion of the dataset, similar to \emph{fixed swapping}.


\section{Conclusions}\label{sec:conclusion}

This work presented LEMURS, an algorithm that learns robot interactions from trajectory demonstrations using self-attention and Hamiltonian-based neural ODE networks. LEMURS advances the state of the art by learning control policies that generalize to increasing numbers of robots and time-varying communications. Our evaluation shows that LEMURS learns behaviors such as collision avoidance and flocking formation from state-only trajectories of few robots, and successfully replicates the tasks in larger robot teams.


\appendices
\section{Network and Experiment Parameters}\label{sec:parameters}

The $\mathsf{SA}$ architecture is parameterized as follows:
\begin{itemize}
    \item $[\mathbf{R}_{\bftheta}]_{ij}$: $W = 3$, $h_w=[4, 8, 8]$, $r_w = [8, 8, 8]$, $d_w = [8, 8, 16]$; functions $\beta = \mathsf{sigmoid}$, $\gamma=\alpha=\mathsf{swish}$~\cite{ramachandran2017searching}. 
    \item $[\mathbf{J}_{\bftheta}]_{ij}$: $W = 3$, $h_w=[4, 8, 8]$, $r_w = [8, 8, 8]$, $d_w = [8, 8, 1]$; functions $\beta = \mathsf{sigmoid}$, $\gamma=\alpha=\mathsf{swish}$~\cite{ramachandran2017searching}; and $[\mathbf{J}_{\bftheta}]_{ij} = \mathbf{0}$ $\forall i \neq j$.
    \item $H^{i}_{\bftheta}$: $W = 3$ layers, $h_w=[6, 8, 8]$, $r_w = [8, 8, 8]$, $d_w = [8, 8, 25]$; functions $\beta = \mathsf{sigmoid}$, $\gamma=\alpha=\mathsf{swish}$~\cite{ramachandran2017searching}. 
\end{itemize}
The network input $\mathbf{X}_i^0$ is an offset version of \eqref{eq:concatenated_neighbors} as follows:
\begin{itemize}
    \item $[\mathbf{R}_{\bftheta}]_{ij}$ and $[\mathbf{J}_{\bftheta}]_{ij}$: $\mathbf{X}_i^0 = \brl{\Delta \bfx_{ii}, \{\Delta \bfx_{ij}\}_{j\in\calN_i^k\backslash\{i\}}}$.
    \item $H^{i}_{\bftheta}$: 
    $$\scaleMathLine{\mathbf{X}_i^0 = \brl{\{\Delta \bfx_{ii}, 0, 0\}, \{\Delta \bfx_{ij}, \Vert \Delta \bfx_{ij}\Vert^{\frac{1}{4}}_2, \Vert\Delta \bfx_{ij}\Vert_2\}_{j\in\calN_i^k\backslash\{i\}}},}$$
    where $\Delta \bfx_{ii} = \bfx_i - \bar{\bfx}_i(KT)$ and $\Delta \bfx_{ij} = \bfx_i - \bfx_j$.
\end{itemize}
The other networks are as follows. For the $\mathsf{MLP}$ and $\mathsf{GNN}$,  $W=1$ and $(4\times n) \times (4\times 4\times n)$ parameters from $1$ unbaised layer/filter; for $H$, $W=1$ and $(6\times n) \times (5\times 5\times n)$  parameters from $1$ unbaised layer/filter. The $\mathsf{GNNSA}$ has $3 \times (4 \times 8) + (8\times 16)$ additional parameters from three self-attention matrices and one self-attention vector.

\balance
\bibliographystyle{IEEEtran}
\bibliography{IEEEabrv,IEEEexample.bib}

\end{document}